\documentclass[preprint, prb, nopacs, superscriptaddress]{revtex4}
\usepackage{epsfig}
\begin{document}
\topmargin-1.0cm

\title {
Theoretical analysis of magnetic coupling in sandwich clusters
V$_n$(C$_6$H$_6$)$_{n+1}$}

\author{Hongming Weng}\email[Author to whom correspondence should be addressed. E-mail:]{hmweng@jaist.ac.jp }
\affiliation {Research Center for Integrated Science, Japan Advanced
Institute of Science and Technology, Nomi, Ishikawa 923-1292, Japan}
\author{Taisuke Ozaki}
\affiliation {Research Center for Integrated Science, Japan Advanced
Institute of Science and Technology, Nomi, Ishikawa 923-1292, Japan}
\author {Kiyoyuki Terakura}
\affiliation {Research Center for Integrated Science, Japan Advanced
Institute of Science and Technology, Nomi, Ishikawa 923-1292, Japan}
\affiliation {Creative Research Initiative 'Sousei', Hokkaido
University, Sapporo 001-0021, Japan} \affiliation {Research
Institute for Computational Sciences, AIST, Tsukuba, Ibaraki
305-8568, Japan}
\date{\today}

\begin{abstract}
The mechanism of ferromagnetism stability in sandwich clusters
V$_n$(C$_6$H$_6$)$_{n+1}$ has been studied by first-principles
calculation and model analysis. It is found that each of the three
types of bonds between V and benzene (Bz) plays different roles. V
3d$_{z^2}$ orbital, extending along the molecular axis, is weakly
hybridized with Bz's HOMO-1 orbital to form the $\sigma$-bond. It is
quite localized and singly occupied, which contributes 1$\mu_B$ to
the magnetic moment but little to the magnetic coupling of
neighboring V magnetic moments. The in-plane d$_{x^2-y^2}$, d$_{xy}$
orbitals are hybridized with the LUMO of Bz and constitute the
$\delta$-bond. This hybridization is medium and crucial to the
magnetic coupling though the $\delta$ states have no net
contribution to the total magnetic moment. d$_{xz}$, d$_{yz}$ and
HOMO of Bz form a quite strong $\pi$-bond to hold the molecular
structure but they are inactive in magnetism because their energy
levels are far away from the Fermi level. Based on the results of
first-principles calculation, we point out that the ferromagnetism
stability is closely related with the mechanism proposed by Kanamori
and Terakura [J. Kanamori and K. Terakura, J. Phys. Soc. Jpn. 70,
1433 (2001)]. However, the presence of edge Bz's in the cluster
introduces an important modification. A simple model is constructed
to explain the essence of the physical picture.
\end{abstract}


\maketitle

%******************************************************************************
\section{introduction} \label{introduction}
Since the discovery of ferrocene Fe(C$_5$H$_5$)$_2$ and
determination of its structure, intensive research of the sandwich
complexes where a metal atom or ion is sandwiched with two organic
molecules (mostly aromatic molecules) has been performed in the
field of organometallic chemistry.\cite{organometallic} The growing
interests in these complexes are mainly due to their potential
usages in catalysis, novel magnetic and optical materials and other
various applications. Among these compounds, the
3d-transition-metal-organic-molecule complexes are particularly
under current active study since molecular magnets are considered as
potential candidates for future applications in high-density
information storage. Therefore, since the discovery of
ferromagnetism in vanadium-benzene V$_n$Bz$_m$ complexes,\cite{mbz,
mbz2} many experimental and theoretical studies\cite{ionization,
wang, jena, bluegel} have been performed in order to shed light on
the mechanism of their ferromagnetism stability. Now, it is
generally accepted that for small size V$_n$Bz$_m$ complexes
(n$\leq$4), the energetically preferred structures are linear
multiple-decker sandwich-like with $m=n-1$, $n$ and $n+1$. The
Stern-Gerlach deflection experiments show that the magnetic moment
in V$_n$Bz$_{n+1}$ increases linearly with $n$.\cite{mbz2} {\it Ab
initio} calculations have confirmed this\cite{wang, jena} though the
magnitude of the measured magnetic moment is smaller than the
theoretical value by about factor two, which is attributed to the
spin relaxation effects during the measurement.\cite{mbz2,
spinrelax} Also, for the ideal infinite (VBz)$_{n=\infty}$ chain,
first-principles calculations predicted that the ferromagnetic (FM)
state is more stable than the anti-ferromagnetic (AFM)
state.\cite{kasai, xiang, prlvbz} A mechanism of double exchange
interaction has been proposed\cite{kasai, xiang, prlvbz} to explain
the ferromagnetism in this complex based on the electronic
structures obtained with the density functional theory calculations
within the local density approximation (LDA) or generalized gradient
approximation (GGA).

However, it is well known that LDA or GGA does not reliably describe
the localized state in transition metal compounds, especially for
these low-dimensional complexes. Including on-site Coulomb
interaction $U$ usually can improve the description and lead to a
more realistic picture. In this paper, by performing GGA+$U$
calculations, we find that the electronic structure is significantly
modified. A mechanism, originally proposed by Kanamori and
Terakura,\cite{kt} is found to be responsible for the stability of
FM state both in infinite and finite V-Bz sandwich complexes. A
similar mechanism was also proposed by Yabushita et al.\cite{mbz,
mbz2, ionization} for the ferromagnetism stability in V$_2$Bz$_3$.
However, it is also found that the presence of edge Bz introduces an
important modification for finite complexes. To show the essence of
the physical picture, a simple model is constructed and the related
parameters are fitted from the first-principles calculations of
$n=2$ cases. The numerical model analysis is semi-quantitatively
consistent with the first-principles calculations for $n$=2, 3 and 4
cases. The model calculation is also used to analyze the size
dependence and configuration dependence of the energy cost for the
magnetic moment reversal.

\section{Methodology} \label{Methodology}
All the first-principles calculations were performed using our
software package OpenMX,\cite{openmx} which is based on a linear
combination of localized pseudo-atomic orbital (LCPAO) method. The
PAOs are generated by a confinement potential scheme\cite{lcpao}
with a cutoff radius of 5.0 a.u. for hydrogen and carbon and 6.5
a.u. for vanadium, respectively. In the pseudopotential generation,
the semicore 3s and 3p states of V atom were included as valence
states. The exchange correlation energy functional within GGA as
parameterized by Perdew, Burke and Ernzerhof (PBE) \cite{pbe} is
used. Double-valence and polarization orbitals were included and the
basis set completeness was checked by optimizing the structure of
single Bz molecule and body centered cubic V metal and comparing
with the experimental values or all electron full-potential
calculations. The real-space grid technique\cite{realspace} was used
with an energy cutoff of 250 Ry in numerical integrations and in the
solution of the Poisson equation using the fast Fourier
transformation algorithm. Hubbard $U$ is included by the approach
proposed in Ref. \onlinecite{plusU} to improve the description of
localized states like V 3d orbitals. All the geometrical structures
of V$_n$Bz$_{n+1}$ complexes are fully optimized under each
individual magnetic configuration until the forces are less than
$1.0\times10^{-4}$ a.u.

\section{results and discussions} \label{result}
\subsection{Ideal infinite (VBz)$_{n=\infty}$ chain}\label{VBzchain}
To understand the mechanism of ferromagnetism stability in V-Bz
multiple-decker sandwich complex, we start by studying the bonding
nature between Bz and V. In general, the molecular axis (z-axis) is
defined as the line that passes through the metal and the center of
gravity of Bz molecule and classify the valence orbitals in terms of
their pseudo-angular momenta around this axis. According to the
symmetry, the six $\pi$-orbitals of benzene form one $p\sigma$
(HOMO-1), two $p\pi$ (HOMO), two $p\delta$ (LUMO) and one $p\phi$
(LUMO+1) orbitals. When sandwiched by two Bz molecules, the five 3d
orbitals of the metal atom are classified to one $d\sigma$
($d_{z^2}$), two $d\pi$ ($d_{xz}$ and $d_{yz}$), and two $d\delta$
($d_{xy}$ and $d_{x^2-y^2}$) orbitals, and the 4s orbital is
classified to an $s\sigma$ orbital.\cite{ionization, prlvbz} To see
the overall feature of the electronic structure, we performed
first-principles calculations for both FM and AFM states of the
ideal infinite V-Bz chain within GGA. The total and partial
densities of states (DOS) for the FM state are plotted in Fig. 1,
which are basically the same as other calculations.\cite{kasai,
xiang, prlvbz} By examining these DOSs and the symmetry of orbital
wave-functions, one can easily find: 1) The coupling between
$p\sigma$ and $d\sigma$ is very weak, nearly non-interacting, making
$d\sigma$ localized and largely spin-split. 2) The $p\pi$ and $d\pi$
orbitals have strong hybridization, making both bonding and
antibonding states far away from the Fermi-level, either totally
occupied or totally empty. They are responsible to the stability of
the molecular structure but not active in magnetic or electronic
properties. 3) The $p\delta$-$d\delta$ hybridization is of medium
strength among the three bonds. $d\delta$ states are also spin split
due to the Hund coupling with the spin polarized $d\sigma$ states. V
atom has 5 electrons, which occupy the $d\sigma$ and doubly
degenerate $d\delta$ bands. Therefore, the number of holes existing
in the overlapping $d\sigma$ and $d\delta$ bands of the minority
spin state is one. Due to this metallic feature, the double-exchange
mechanism has been proposed to be the possible origin of
ferromagnetism.\cite{kasai, prlvbz, xiang} The FM state is more
stable than the AFM state by 0.091 eV per VBz (see Table I). The
total magnetic moment is just 1 $\mu_B$ per VBz unit with its
integer value being consistent with the half-metallic feature of
DOS. In fact, the local magnetic moment on V atoms is about 1.15
$\mu_B$. The total magnetic moment is reduced to 1 $\mu_B$ due to
the negative magnetic moment on Bz being polarized through the
$p\delta$-$d\delta$ hybridization.\cite{prlvbz}

\begin{figure}
\centering
\includegraphics[width=0.8\textwidth]{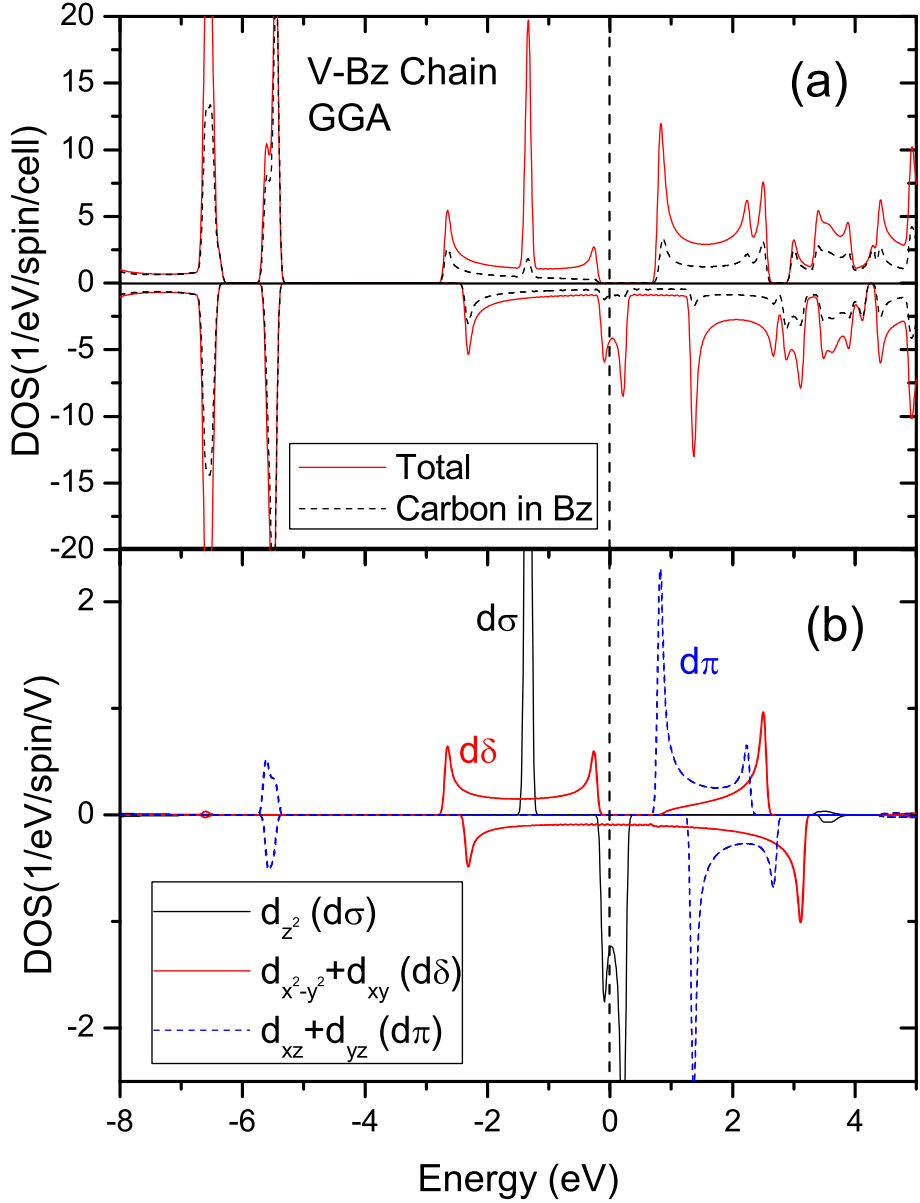}
\caption{(Color online) DOS of a FM infinite V-Bz chain calculated
within GGA. (a) Total DOS and the partial DOS of carbons in Bz
molecules; (b) partial DOS for each of V 3d orbitals.}\label{fig1}
\end{figure}

\begin{figure}
\centering
\includegraphics[width=0.8\textwidth]{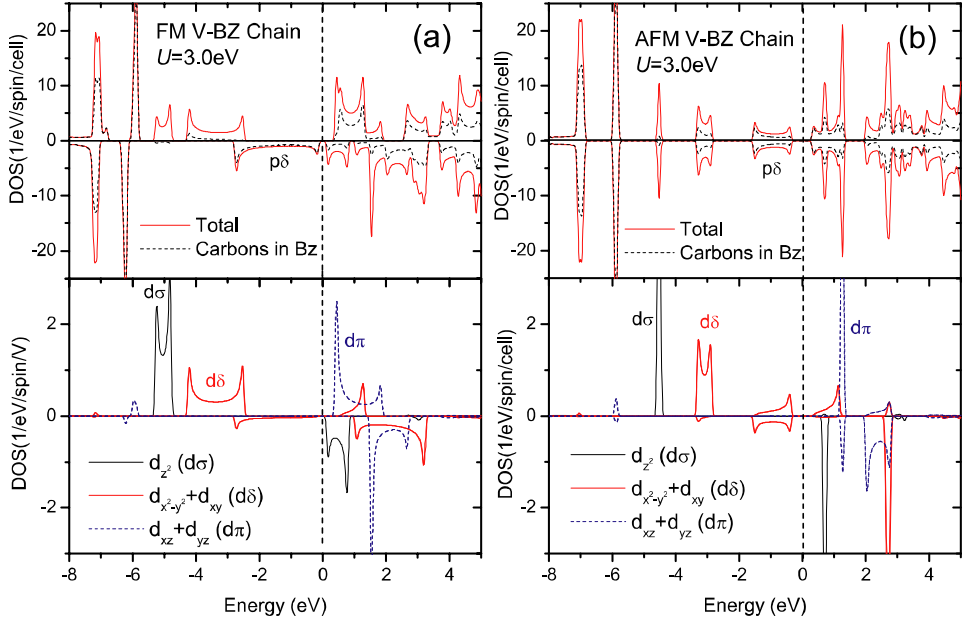}
\caption{(Color online) DOS of an infinite V-Bz chain calculated
within GGA+$U$ ($U$=3.0 eV) for (a) FM and (b) AFM state. Upper
panel of them is total and the partial DOS of carbons in Bz
molecules, while the lower panel is partial DOS for each of V 3d
orbitals.}\label{fig2}
\end{figure}

\begin{figure}
\centering
\includegraphics[width=0.8\textwidth]{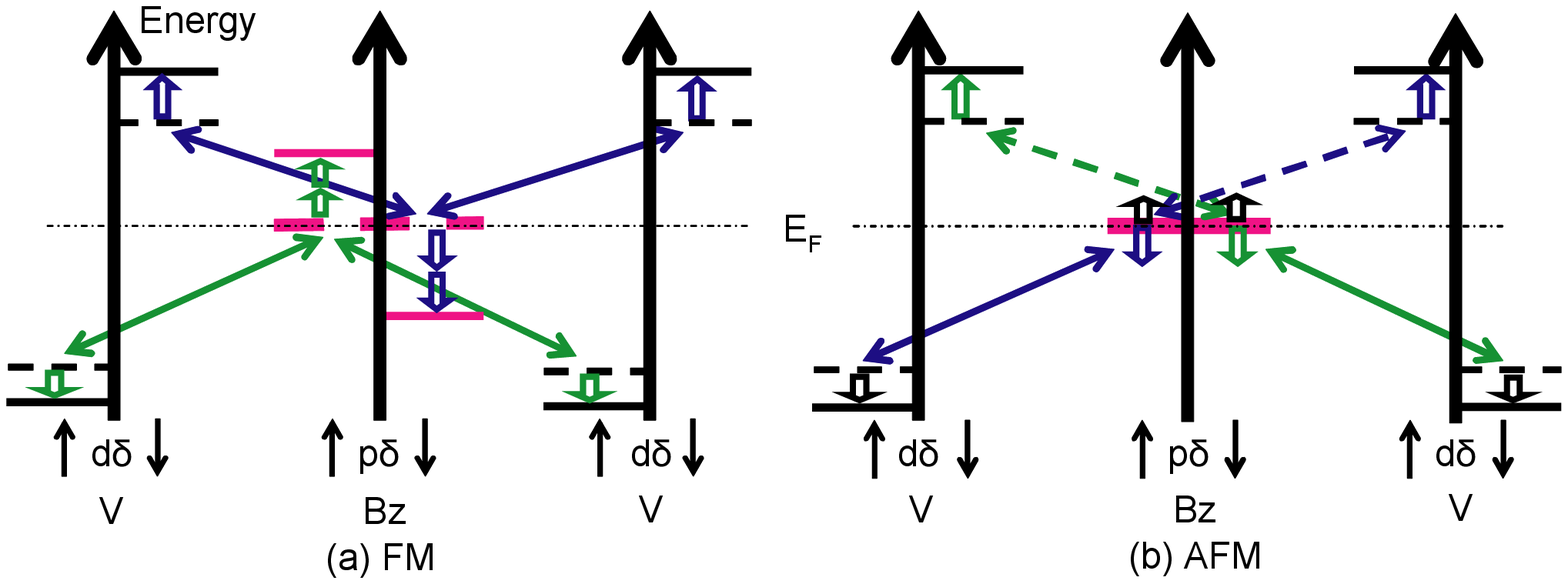}
\caption{(Color online) A schematic picture describing the
interactions between V $d\delta$ and intervening Bz $p\delta$ states
in FM and AFM configurations. The thick vertical up arrow indicates
the energy increasing direction and its left (right) side represents
the majority- (minority-) spin channel. The horizontal solid
(dashed) lines represent the energy levels of the corresponding
state after (before) $p-d$ hybridization. The hollow arrows
indicates the direction and magnitude of movements for each state.
Fermi level is labeled by the horizontal dash-dotted
line.}\label{fig3}
\end{figure}

However, for localized states like $d\sigma$ and $d\delta$, LDA or
GGA is not generally reliable. GGA+$U$ could improve the description
of the electronic structure. In the present calculation, $U$ is a
parameter but we have confirmed that the qualitative features are
not affected by changing $U$ from 2.0 eV to 4.0 eV. As a typical
case, the results with $U$=3.0 eV will be mostly shown below. The
DOS of GGA+$U$ calculation with $U$=3.0 eV is shown in Fig.2a for
the FM state. As expected, the spin-splitting in $d\sigma$ and
$d\delta$ is largely enhanced compared with that in GGA calculation.
It is interesting to note that the broad band in the minority-spin
state right below the Fermi level is mostly composed of the
$p\delta$ states, which are LUMO in isolated Bz. Therefore,two
$d\delta$ electrons of V are formally transferred to $p\delta$
states of Bz. The spin splitting in $d\delta$ states is now larger
than the energy separation of $d\delta$ and $p\delta$ levels before
the onset of the $p-d$ hybridization. (More detailed discussion will
be given below.) A similar calculation for the AFM state gives the
DOS shown in Fig.2b. Now, both the FM and AFM states are insulating
and yet the stability of the FM state against the AFM state is
enhanced (see Table I).  This implies that the mechanism of the
ferromagnetism stability is not the double exchange.  In the
following, we further analyze the electronic structure in detail and
elucidate the mechanism of the magnetic coupling between V magnetic
moments.

As $d\sigma - p\sigma$ hybridization is weak, $d\sigma$ states does
not make significant contribution to the magnetic coupling.
Moreover, even if there is any contribution from them it will be AFM
superexchange. The magnetic coupling is mostly governed by the
$d\delta$ and $p\delta$ states. Therefore, we will focus only on the
$\delta$ states. Figure 3 is a schematic diagram showing the basic
feature of $p-d$ hybridization in the $\delta$ states. Due to the
strong exchange splitting of $d\delta$ states, their majority-spin
states are below the $p\delta$ states and fully occupied. On the
other hand, the minority-spin $d\delta$ states are above the
$p\delta$ states and empty. In the FM configuration of the V
magnetic moments, the majority-spin $p\delta$ states are strongly
pushed up in energy by the $p-d$ hybridization and {\it vice versa}
in the minority-spin state. In the AFM configuration, on the other
hand, the energy shift of $p\delta$ states is much reduced because
the hybridization with the two neighboring $d\delta$ states produces
energy shifts with opposite sign. In addition to this, the energy
shift is the same in both spin states. One additional comment is
that the apparent exchange splitting of $d\delta$ states is enhanced
by the $p-d$ hybridization in both the FM and AFM states. Based on
the observation of the energy shift scheme, we show a schematic DOS
for $\delta$ states in the FM state in Fig. 4. In the absence of
$p-d$ hybridization, the doubly degenerate $p\delta$ bands are half
filled (Fig.4a).  With the $p-d$ hybridization which produces large
energy shifts in the $p\delta$ states (Fig.3a), the DOS is modified
to the one shown in Fig.4b, which captures the essential feature of
the $\delta$ symmetry part of the DOS in Fig.2a. Now, the electrons
originally occupying the majority-spin $p\delta$ bands in Fig.4a are
transferred to the minority-spin $p\delta$ bands as shown in Fig.4b,
which produces a large energy gain in the FM state. A large negative
spin polarization on Bz is also naturally explained. This is exactly
the same story proposed by Kanamori and Terakura.\cite{kt} The
importance of the spin polarization in the intervening Bz for the
ferromagnetism stability was also pointed out by Yabushita et al.
for V$_2$Bz$_3$ cluster.\cite{mbz, mbz2, ionization}

As shown in Table I, the total energy difference between FM and AFM
states increases as $U$ value increases. Although the local magnetic
moment on V ion increases with $U$, the negative polarization on Bz
also increases and keeps the total magnetic moment per VBz unit the
same as 1.0 $\mu_B$, which is produced by $d\sigma$ states.

\begin{table}[htbp]
\caption{ Magnetic moment on V ($\mu_B$) and total energy difference
between FM and AFM states ($\Delta $E= E$_{FM}$-E$_{AFM}$)
calculated with different $U$ values. }
\begin{tabular}
{|c|c|c|c|c|} \hline \raisebox{-1.50ex}[0cm][0cm]{U (eV)}&
\multicolumn{2}{|c|}{Infinite V-Bz Chain} &
\multicolumn{2}{|c|}{V$_2$Bz$_3$ }  \\
\cline{2-5}
 &
Magnetic moment on V & $\Delta $E/V (eV)& Magnetic moment on V&
$\Delta $E/cluster (eV) \\
\hline 0.0& 1.15& -0.091& 1.10 & -0.003 \\
\hline 1.0& 1.35& -0.113& 1.17& -0.004 \\
\hline 2.0& 1.67& -0.109& 1.29& -0.004 \\
\hline 3.0& 2.03& -0.184& 1.50& -0.005 \\
\hline 4.0& 2.23& -0.251& 1.82& -0.007 \\
\hline
\end{tabular}
\label{tab1}
\end{table}

\begin{figure}
\centering
\includegraphics[width=0.8\textwidth]{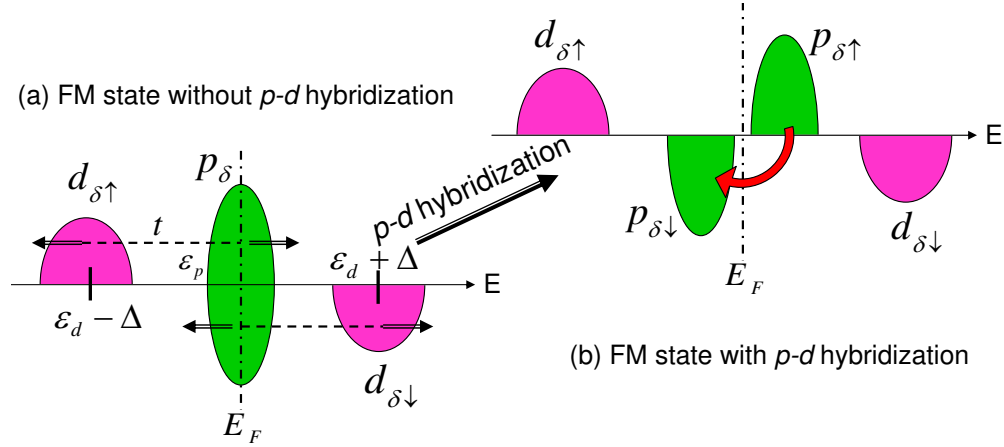}
\caption{(Color online) Schematic pictures describe the mechanism in
which the $p-d$ hybridization stabilize the FM states. The DOS (a)
without and (b) with $p-d$ hybridization is plotted. The energy gain
due to the charge transfer is indicated.}\label{fig4}
\end{figure}

\subsection{Finite V$_n$Bz$_{n+1}$ clusters}\label{VBzcluster}
Now, we come to the study of finite size sandwich clusters
V$_n$Bz$_{n+1}$. For $n$=1, the spin-polarized VBz$_2$ with S=1/2
has lower energy than the non-magnetic one, which is consistent with
the experimental measurement. For $n$=2, as shown in Table I the
negative total energy difference between FM and AFM states indicates
the FM state is more stable, being also consistent with experimental
observations. However, the value of the energy difference is only
several meV, much smaller than that in the infinite chain. Other
calculations also give the energy difference less than 10
meV.\cite{mbz2,wang} Before presenting further detailed analysis,
the following comment on Table I may be needed for the comparison of
energies. The total energy difference is for one VBz unit in the
infinite chain, while it is for V$_2$Bz$_3$ in the cluster. A
justification of this seemingly strange comparison comes from an
expectation that the magnetic coupling comes mostly from the energy
change associated with the spin polarization of the intervening Bz
between the V atoms.  In this sense, both units have only one Bz
shared by two neighboring V atoms.

According to the physical picture described in the infinite chain,
the magnetic coupling in V$_2$Bz$_3$ can be studied by a simple tight-binding
model which takes account of the hybridization between $p\delta$ and $d\delta$ states.
For example the model Hamiltonian matrix for the AFM state is given as
\[
\left| {{\begin{array}{*{20}c}
 {\varepsilon _p } & t'                           &         0         & 0                          & 0  \\
     t'             & {\varepsilon _d - \Delta }  &         t         & 0                          & 0  \\
     0             & t                           & {\varepsilon _p } & t                          & 0 \\
     0             & 0                           & t                 & {\varepsilon _d + \Delta } & t' \\
     0             & 0                           & 0                 & t'                          &{\varepsilon _p } \\
\end{array} }} \right|.
\]
where $\epsilon_p$ and $\epsilon_d$ are the on-site energies of
$p\delta$ and $d\delta$ levels, respectively, and 2$\Delta$ is the
exchange splitting. In order to see the role of the edge Bz
explicitly, the hopping integral between $d\delta$ and $p\delta$ is
expressed as $t'$ for the edge Bz part and $t$ for the middle Bz
part.

To remove the edge Bz's, we adopt a cyclic boundary condition in the
(VBz)$_2$ cluster. The Hamiltonian, which is identical to the $\Gamma$
point expression of an infinite chain, is written as
\[
\left| {{\begin{array}{*{20}c}
 {\varepsilon _p }  &             t              &   0    &    t  \\
          t         & {\varepsilon _d - \Delta } &   t    &    0  \\
          0         &             t              & {\varepsilon _p }  & t  \\
          t         &             0              &    t   & {\varepsilon _d + \Delta }   \\
\end{array} }} \right|.
\]
The eigenvalues of these Hamiltonians can be calculated easily.
Assuming $\epsilon_p$=$\epsilon_d$=$\epsilon_0$ and taking the
approximation that $|t/\Delta| << 1.0$ , we show the energy diagrams
in Fig. 5. In the $n=2$ case, there are 10 electrons introduced by
two V atoms. Two of them occupy the $d\sigma$ states and the other 8
electrons should occupy the $d\delta$ and $p\delta$ states depicted
here. Note that both $d\delta$ and $p\delta$ states have double
orbital degeneracy. Fermi level can be determined as shown in the
energy diagrams. In order to understand the overall features of
energetics among four different configurations shown in Fig.5, we
assume $t'=t$, which is reasonable since the V-Bz distance is almost
equally spaced. Then we note the following aspects in the energy
diagram.

1) As for the energy levels originating from $d\delta$ states, the
sum of the occupied part of orbital energies is the same for all
cases shown in Fig.5.  Note that the spin degeneracy has to be taken
into account in the case of AFM state. Therefore, the total energy
difference among the four cases shown in Fig.5 comes entirely from
the energy diagram associated with $p\delta$ states.

2) In the case of FM state, even for the energy levels originating
from $p\delta$ states, the sum of occupied part of orbital energies
is the same for both isolated cluster and cyclic boundary condition.

3) In the case of AFM state with cyclic boundary condition, as
there is no shift of energy levels of $p\delta$ states, the FM state
is definitely lower in energy by $8t^2/ \Delta$ per cluster.

4) In the case of AFM state for isolated cluster, the energy of one
of the $p\delta$ states is lowered due to the spin polarization at
the edge Bz.  Therefore the edge Bz contributes to the stabilization
of AFM state of the isolated cluster. The FM state is still lower in
energy than the AFM state but now only by $4(2-\sqrt{3})t^2/ \Delta$
per cluster.  This is basically the reason for the significant
reduction in the stability energy of the FM state against the AFM
state in the isolated cluster as shown in Table I.

Now we give some additional comments on the role of edge Bz in the
stabilization of AFM state in isolated cluster. If there is no
electronic coupling among the Bz's, the energy level shift of the
edge $p\delta$ state will be simply $-{t'}^2/ \Delta$. Then the
stability energy of the FM state against the AFM state for the
isolated cluster will become $4t^2/ \Delta$, just half of that for
the cyclic boundary condition, as is naively expected.  However,
because the $p\delta$ states of the three Bz's form a coupled
degenerate system, all the three $p\delta$ states contribute to the
polarization of the edge Bz and the actual energy shift is much
larger as shown in Fig.5. This can be seen in the AFM DOS of Fig. 6.
The state around -1.0 eV is the magnetically polarized Bz $p\delta$
state which consists mainly of the edge Bz state and also partly of
the middle Bz state. Although the FM state is still more stable than
the AFM state for any values of $t'$ and $t$, the energy difference
is reduced to $2(t/t')^2 (t^2/ \Delta)$ when $|t/t'|<<1.0$.

\begin{figure}
\centering
\includegraphics[width=0.8\textwidth]{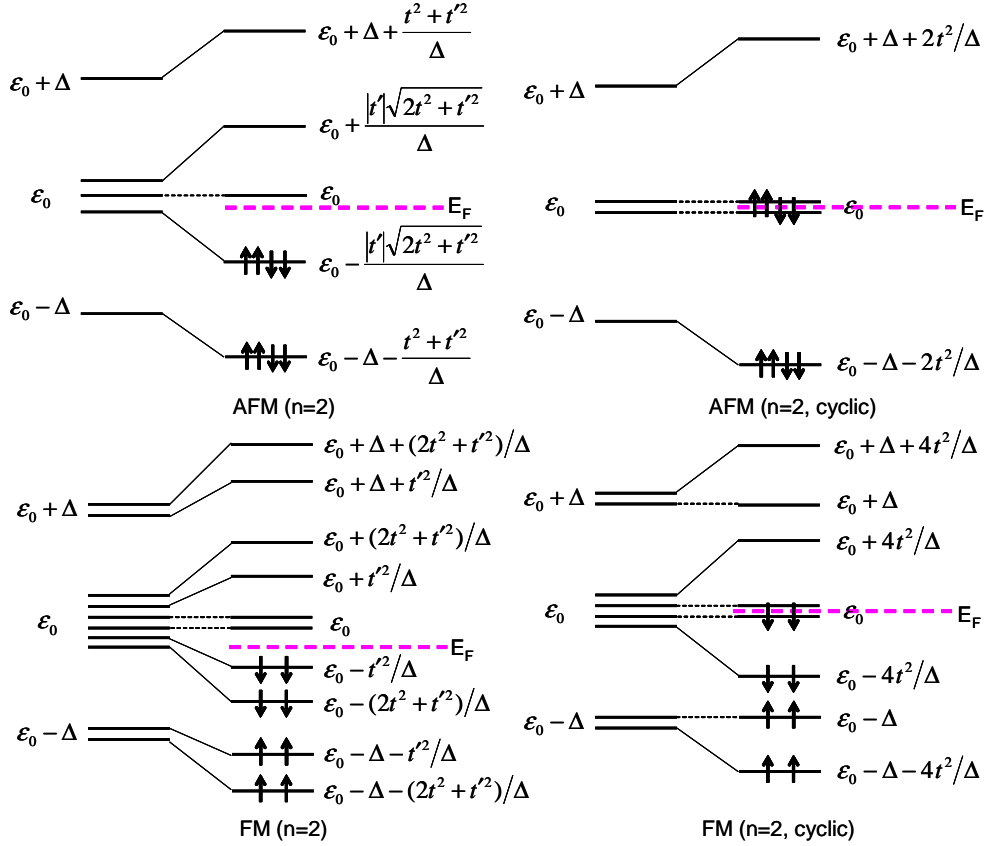}
\caption{(Color online) Energy diagrams for AFM and FM V$_2$Bz$_3$
cluster and (VBz)$_2$ with cyclic boundary condition. $\epsilon_{p}=
\epsilon_{d}=\epsilon_{0}$ is assumed and the energy levels are
given only in the lowest order of $t^2/\Delta$ and
$t'^2/\Delta$.}\label{fig5}
\end{figure}

\begin{figure}
\centering
\includegraphics[width=0.8\textwidth]{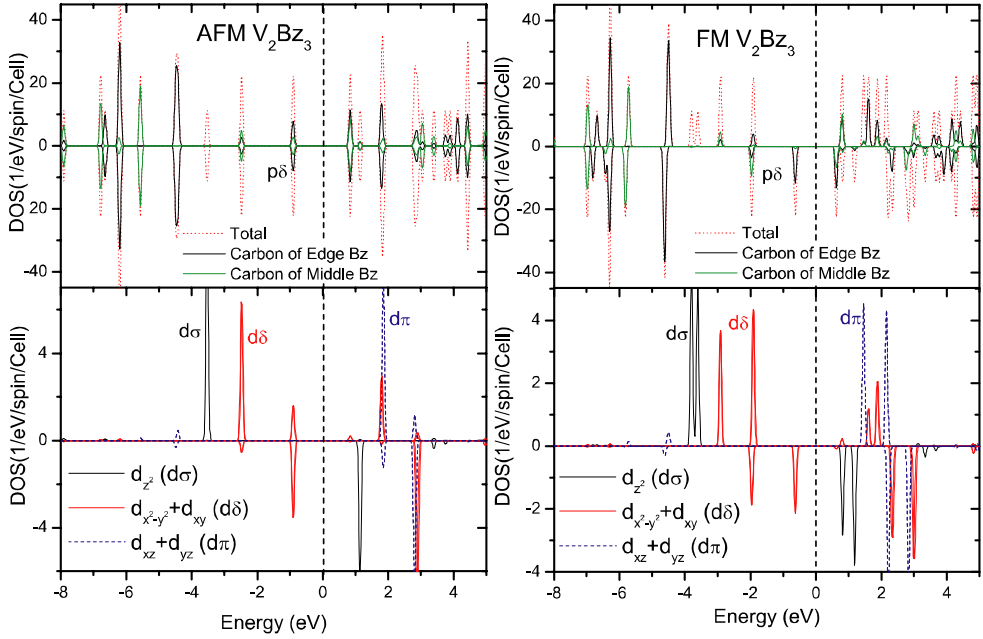}
\caption{(Color online) Total and partial DOS for AFM and FM
V$_2$Bz$_3$ cluster calculated with GGA+$U$ and $U$=3.0
eV.}\label{fig6}
\end{figure}

\begin{table}[htbp]
\caption{Magnetic configuration (Arrow indicates the direction of
local moment on V atoms and vertical lines represent Bz's) of
V$_3$Bz$_4$, number of Bz's sandwiched with antiparallel V magnetic
moments (represented by vertical dotted lines) and the energy
difference (in eV) relative to the FM state from model calculation.
Those in parentheses are from first-principles GGA+$U$ calculations
with $U$=3.0 eV. The structure is optimized for each magnetic
configuration in the GGA+$U$ calculations, while $t=t'=$1.648 eV is
used in the model analysis.}
\begin{tabular}
{|c|c|c|} \hline magnetic configuration & No. of $\uparrow \vdots
\downarrow$ pairs & Energy difference relative to FM state \\
\hline $\arrowvert \uparrow \arrowvert \uparrow \arrowvert \uparrow \arrowvert$ & 0 & 0.0 \\
\hline $\arrowvert \uparrow \arrowvert \uparrow \vdots \downarrow \arrowvert$ & 1& 0.033 (0.018) \\
\hline $\arrowvert \uparrow \vdots \downarrow \vdots \uparrow \arrowvert$ & 2& 0.060 (0.037) \\
\hline
\end{tabular}
\label{tab2}
\end{table}

\begin{table}[htbp]
\caption{Magnetic configurations (Arrow indicates the direction of
local moment on V atoms and vertical lines represent Bz's) of
V$_4$Bz$_5$, number of Bz's sandwiched with antiparallel V magnetic
moments (represented by vertical dotted lines) and the energy
difference (in eV) relative to the FM state from model calculation.
Those in parentheses are from first-principles GGA+$U$ calculations
with $U$=3.0 eV. The structure is optimized for each magnetic
configuration in the GGA+$U$ calculations, while $t=t'=$1.648 eV is
used in the model analysis.}
\begin{tabular}
{|c|c|c|} \hline magnetic configuration & No. of $\uparrow \vdots
\downarrow$ pairs & Energy difference relative to FM state \\
\hline $\arrowvert \uparrow \arrowvert \uparrow \arrowvert \uparrow \arrowvert \uparrow \arrowvert $ & 0 & 0.0 \\
\hline $\arrowvert \uparrow \arrowvert \uparrow \arrowvert \uparrow \vdots \downarrow \arrowvert$ & 1& 0.037 (0.036)\\
\hline $\arrowvert \uparrow \arrowvert \uparrow  \vdots \downarrow \arrowvert \downarrow \arrowvert$ & 1& 0.051 (0.056)\\
\hline $\arrowvert \uparrow \vdots \downarrow  \arrowvert \downarrow \vdots \uparrow \arrowvert$ & 2& 0.073 (0.065) \\
\hline $\arrowvert \uparrow \arrowvert \uparrow  \vdots \downarrow \vdots \uparrow \arrowvert$ & 2& 0.077 (0.082) \\
\hline $\arrowvert \uparrow \vdots \downarrow  \vdots \uparrow \vdots \downarrow \arrowvert$ & 3& 0.105 (0.103)\\
\hline
\end{tabular}
\label{tab3}
\end{table}

\begin{figure}
\centering
\includegraphics[width=0.8\textwidth]{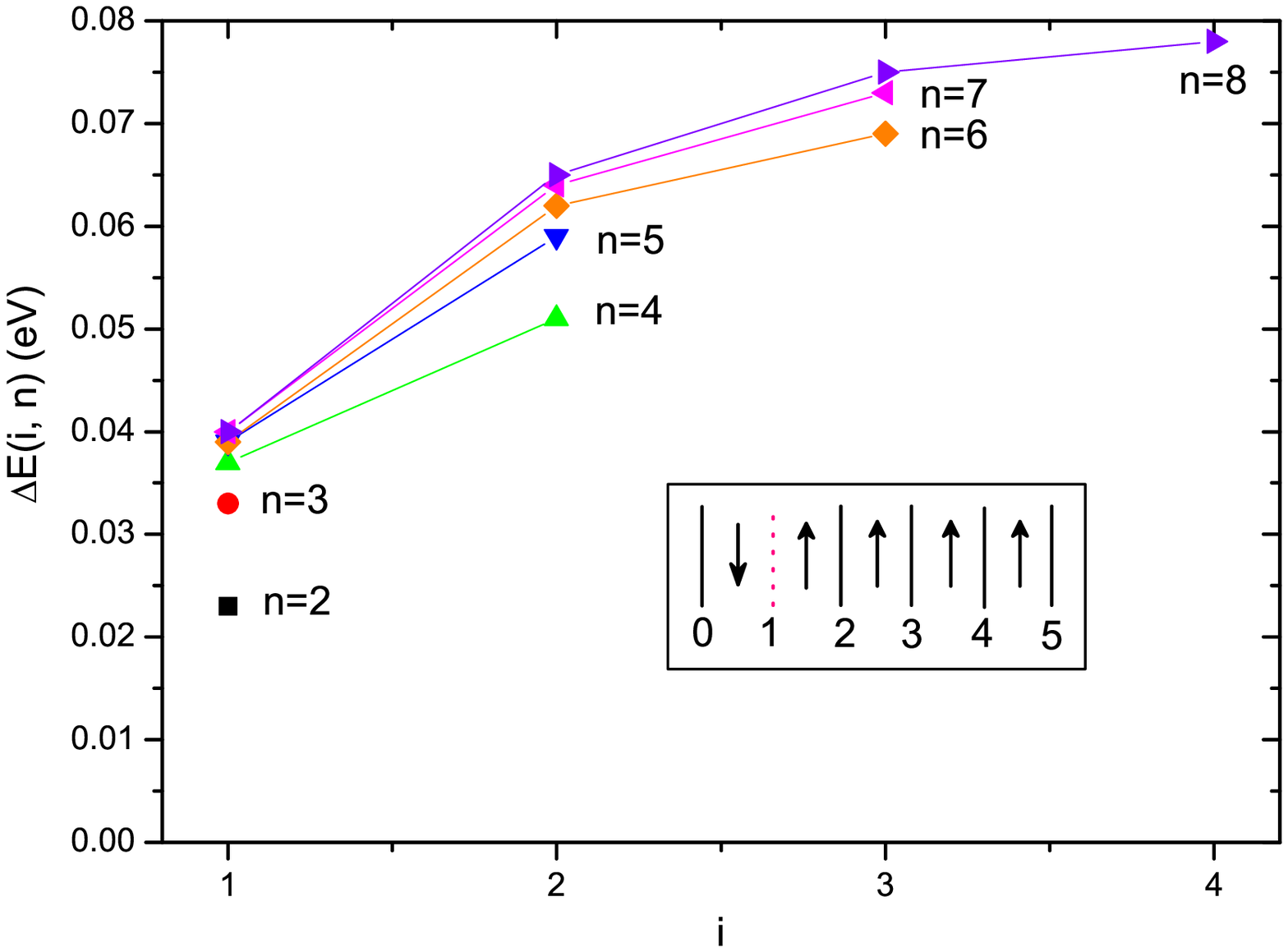}
\caption{(Color online) Site $i$ and cluster size $n$ dependence of
the energy cost $\Delta E(i,n)$ to have an antiparallel magnetic
moment pair. The site $i$ corresponds to the location of the Bz
sandwiched with antiparallel magnetic moments of V and the cluster
size $n$ corresponds to V$_n$Bz$_{n+1}$. $t=t'$ is assumed. In the
inset, the configuration is specified with $i$=1 and $n$=5.
}\label{fig7}
\end{figure}

In order to make the model analysis more quantitative, the
parameters in the model are adjusted to reproduce the energy diagram
obtained by the first-principles calculation. The total and partial
DOS for the AFM and FM V$_2$Bz$_3$ calculated with $U$=3.0 eV are
plotted in Fig. 6. Without losing generality, $\epsilon_p$ can be
taken as 0.0 to be the energy reference and $t'$ is first assumed to
be equal to $t$. $t$, $\epsilon_d$ and $\Delta$ are adjusted using
the occupied four levels in the FM state and the values for the
parameters are $t$=1.648 eV, $\epsilon_d=-0.650$ eV and
$\Delta$=0.916 eV, respectively. Then in order to clarify the
effects of variation of $t'/t$ in the relative stability between FM
and AFM states, we assume the following distance dependence of the
$p-d$ hybridization, $t,t' \propto d^{-4}$ with $d$ the distance
between C and V and readjust $t$ and $t'$ keeping $\epsilon_d$,
$\Delta$ and the total energy of the FM state unchanged. According
to our structural optimization for $U$=3.0 eV, the V-C distances are
2.304 \AA$ $ for the middle Bz and 2.248 \AA$ $ for the edge Bz. The
final values of $t$ and $t'$ are 1.555 eV and 1.716 eV,
respectively. By using these parameters, we find that FM V$_2$Bz$_3$
is lower than the AFM one by 0.018 eV. The corresponding value for
the case of $t=t'$ is 0.023 eV. For the cyclic case, FM state is
more stable by 0.245 eV. These results are qualitatively consistent
with our first-principles calculations. If a small contribution from
$d\sigma$ states to AFM coupling may be taken into account,
agreement between the model and the first-principles calculation may
become better.

The model analysis can be easily extended to larger clusters to
study the size effect. Here we show some results for $n$=3 and 4
cases. In larger clusters there are several magnetic configurations.
Table II and Table III summarize the energy cost needed to reverse
the spin magnetic moment of V from the ground state for $n$=3 and 4,
respectively. In the left-end column of the tables, arrows indicate
the direction of V magnetic moment, while the vertical line
corresponds to Bz.  Those Bz's sandwiched with antiparallel V
magnetic moments are denoted by dotted vertical lines. First, we
note that in the GGA+$U$ scheme, the ground state is FM for both
$n=$3 and 4 being consistent with other GGA-BLYP\cite{blyp}
calculations,\cite{wang} though one of the GGA-PBE calculations
claims that the AFM state is the ground state for
$n=$3.\cite{bluegel} Second, the energy cost for reversing the
magnetic moment of one of the edge V's increases with $n$.  This
trend is seen in both the GGA+$U$ calculation and the model
analysis. Third, the energy cost is basically controlled by the
number of dotted vertical lines , i.e., the number of antiparallel
pairs, in the magnetic configuration. Fourth, between the two
configurations with one dotted vertical line (the 2nd and 3rd
configurations in Table III), the one with antiparallel spin
configuration at the cluster edge costs less energy. The second and
the fourth aspects are clearly seen in Fig. 7 where the results by
model calculations are also shown for larger clusters.

%**********************************************************************************
\section{Conclusion} \label{Conclusion}
By examining the electronic structures of the infinite chain of VBz
obtained with the GGA+$U$ calculations, we first found that the
orbitals with different symmetries play distinct roles in the
physical properties of the VBz system. The $d\sigma$ orbital of V,
which has very weak hybridization with $p\sigma$ orbital of Bz, is
localized and singly occupied with a large exchange splitting. It
contributes one $\mu_B$ magnetic moment to the whole system but
plays minor role in the magnetic coupling between neighboring V
local moments. Perhaps, it may contribute to weak anti-ferromagnetic
superexchange. The $d\pi$ and $p\pi$ states hybridize strongly to
push the $\pi$ molecular orbitals far away from the Fermi level
either fully occupied or empty.  Therefore, although they contribute
to the stability of the multiple-decker sandwich structure, they are
inactive in magnetic or electronic properties. The hybridization
between $d\delta$ and $p\delta$ states is medium among the three
types of bonds and these states are crucially important in the
magnetic coupling. We pointed out that the mechanism of
ferromagnetism stability proposed by Kanamori and Terakura\cite{kt}
is directly applicable to the ferromagnetism of the infinite chain
of VBz. We have also found that the presence of edge Bz in
V$_2$Bz$_3$ cluster introduces an interesting modification to the
mechanism. The edge Bz's are not simply inactive in the magnetic
coupling but tend to favor the antiferromagnetic coupling of V
magnetic moments. We set up a simple tight-binding model to analyze
the problem in detail.  Finally we studied the size dependence of
magnetic interaction in the V$_n$Bz$_{n+1}$ clusters.

%**********************************************************************************

\begin{acknowledgments}
We thank Prof. H. Fukuyama and Prof. A. Nakajima for valuable comments and
discussion. This work is partly supported by the Next Generation Supercomputing
Project, Nanoscience Program and also partly by Grant-in-Aids for Scientific
Research in Priority Area "Anomalous Quantum Materials", both from MEXT, Japan.

\end{acknowledgments}

%***************************************************************************
%***************************************************************************

%***************************************************************************

\end{document}